\documentclass[prl,twocolumn,english,reprint, longbibliography, superscriptaddress, breaklinks=true, showkeys, showpacs=false, nofootinbib]{revtex4-1}

\usepackage[T1]{fontenc}
\usepackage[utf8]{inputenc}
\usepackage{color}
\usepackage{babel}
\usepackage{amsmath}
\usepackage{amssymb}
\usepackage{subfigure}
\usepackage{graphicx}
\usepackage{physics} 
\usepackage{dsfont}
\usepackage{hyperref}

\begin{document}

\title{Quantum detailed fluctuation theorem in curved spacetimes: \\ The observer dependent nature of entropy production}

\author{Marcos L. W. Basso}
\affiliation{Center for Natural and Human Sciences, Federal University of ABC, Santo Andr\'e, SP, 09210-580, Brazil}

\author{Jonas Maziero}
\affiliation{Physics Department, Center for Natural and Exact Sciences, Federal University of Santa Maria, Roraima Avenue 1000, Santa Maria, RS, 97105-900, Brazil}

\author{Lucas C. C\'eleri\href{https://orcid.org/0000-0001-5120-8176}{\includegraphics[scale=0.05]{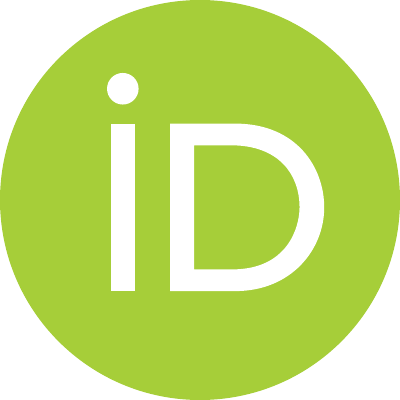}}}
\email{lucas@qpequi.com}
\affiliation{QPequi Group, Institute of Physics, Federal University of Goi\'as, Goi\^ania, GO, 74.690-900, Brazil}

\begin{abstract}
The interplay between thermodynamics, general relativity and quantum mechanics has long intrigued researchers. Recently, important advances have been obtained in thermodynamics, mainly regarding its application to the quantum domain through fluctuation theorems. In this letter, we apply Fermi normal coordinates to report a fully general relativistic detailed quantum fluctuation theorem based on the two-point measurement scheme. We demonstrate how the spacetime curvature can produce entropy in a localized quantum system moving in a general spacetime. The example of a quantum harmonic oscillator living in an expanding universe is presented. This result implies that entropy production is strongly observer dependent and deeply connects the arrow of time with the causal structure of the spacetime.
\end{abstract}


\maketitle
 
\textit{Introduction.} Although the fundamental laws of Nature respect the time-reverse symmetry, irreversible processes are everywhere in the natural world~\cite{zeh}. Irreversibility, marked by the production of thermodynamic entropy~\cite{landi}, establishes the thermodynamic arrow of time, pointing from low to high entropy~\cite{parrondo}. A significant step in this research area has been the formulation of fluctuation theorems, which extend the second law of thermodynamics. These theorems assert that the likelihood of observing a negative entropy production, or a reversal of the arrow of time, vanishes exponentially~\cite{Crooks1998,Jarzynski2011,Seifert2012,Esposito2009, Campisi}. One implication of these findings is that, on average, a positive entropy production will be typically manifested in physical processes.

A cloudy direction is the intersection of relativity and thermodynamics, that dates back to over a century ago, with Einstein and Planck diving into how thermodynamic properties behave under changes in reference frames~\cite{einstein1907,einstein1989,planck1908}. A remarkable advance in this intersection involves the development of black hole thermodynamics~\cite{Wald1994}, which was subsequently used to demonstrate that Einstein's field equations can be interpreted as a thermodynamic equation of state~\cite{Jacobson1995}. This approach has been further applied to investigate the non-equilibrium properties of the spacetime~\cite{Eling2006}. Moreover, attempts for building a statistical mechanical theory of the gravitational field, alongside the suggestion that time may have a thermodynamic origin, were reported in Refs.~\cite{Rovelli1993, Connes1994, Rovelli2011, Rovelli2013}.

Here, by working in this border, we demonstrate how the spacetime curvature can produce entropy in a localized quantum system living in a general spacetime. Some developments in this direction have been achieved. In the realm of linear effects, Mottola~\cite{Mottola1986} established a fluctuation-dissipation relation in curved spacetime. Iso \textit{et al.}~\cite{Iso2011} explored the non-equilibrium fluctuations of a black hole horizon through the application of the Jarzynski equality~\cite{Jarzynski1997} along with the generalized second law of thermodynamics~\cite{Bekenstein1974}. Moreover, a fluctuation theorem for a quantum field in a specific model of an expanding Universe was described in Ref.~\cite{Liu}.

Our central result goes beyond these studies, by proving a fully general relativistic quantum fluctuation theorem, based on the two point measurement (TPM) scheme~\cite{Esposito2009}, for a localized quantum system. We thus extend our findings in Ref.~\cite{Basso2023} in two significant ways. Firstly, the Tasaki-Crooks theorem entails the Jarzynski equality when integrated across the probability distribution. Secondly, and most notably, it unveils the complete impact of the gravitational field on irreversible processes by explicitly considering the effect of spacetime curvature. Our primary finding is illustrated in the case of a quantum mechanical harmonic oscillator living in an expanding universe.

We use natural units throughout the paper. The signature of the metric is $(-,+,+,+)$ and $\eta_{a b} = \text{diag}(-1,1,1,1)$ is the Minkowski metric.

\textit{Fermi normal coordinates.} We consider the stochastic thermodynamics of a non-relativistic quantum system that is localized around a time-like trajectory in an arbitrary spacetime. In order to define physical quantities, we employ Fermi normal coordinates~\cite{Poisson2011}. This choice is based on the fact that such coordinates properly define rest spaces where the Hilbert space can be constructed and all the relevant quantities can be unambiguously defined. Also, using the Fermi transport, it is possible to follow the evolution of the system from one of the rest space to the others.

Our strategy is the following. We first build the Fermi normal coordinates around a time-like trajectory that describes the worldline of our laboratory frame. Then, we consider the Hamiltonian formulation of the dynamics of a localized quantum particle around this time-like trajectory~\cite{Perche2022,Zych2011,Pikovski2015}. This will provide the necessary description of a localized quantum system in a curve spacetime, which is the basic ingredient employed here. 

Technically, we consider a four-dimensional spacetime $(\mathcal{M}, \mathbf{g})$, with $\mathcal{M}$ being a differentiable manifold while $\mathbf{g}$ stands for a Lorentzian metric. The world-line of the laboratory frame is a time-like curve $\gamma: I \subset \mathbb{R} \to \mathcal{M}$, which can be parameterized by its proper time $\tau \in I$. The frame $4$-velocity $u^{\mu}$ fulfils $u_{\mu} u^{\mu} = - 1$. 

The next step is to build the Fermi normal coordinates (see Sec.~I of the Supplemental Material for details). First, we define $\tau$ as the time component of such coordinate system. An orthonormal basis $e_{a}^{\ \mu}$ is then defined at a point $\mathfrak{p} \equiv \gamma(\tau = 0) \in \mathcal{M}$, where $a = 0, 1, 2, 3$ labels the four basis vector while $e_0^{\ \mu}$ is identified with the tangent vector $u^{\mu}$. Thus, at point $\mathfrak{p}$ we have $g_{\mu \nu} e_{a}^{\ \mu} e_{b}^{\ \nu} =  \eta_{a b}$. Now we extend this frame along the trajectory $\gamma$ such that the basis remains orthogonal. This is achieved by transporting the vectors $e_{a}^{\ \mu}$ via the Fermi-Walker transport~\cite{hawkingellis}.

By considering the normal neighborhood $\mathcal{U}_{\mathfrak{p}}$ of the point $\mathfrak{p} = \gamma(\tau = 0)$ we can construct the space-like Fermi normal coordinates. From this, we define the local rest space $\mathcal{R}_{\mathfrak{p}} \subset \mathcal{U}_{\mathfrak{p}}$ of the point $\mathfrak{p}$ as the set of points spanned by all geodesics that start from $\mathfrak{p}$ with tangent vector orthogonal to $u^{\mu}$. Therefore, the coordinates $(\tau = 0, x^1, x^2, x^3)$ can be assigned to any point $\mathfrak{r} \in \mathcal{R}_{\mathfrak{p}}$. Finally, the local rest space of the curve $\gamma$ is defined as $\mathcal{R} \equiv \cup_{\mathfrak{p} \in \gamma} \mathcal{R}_{\mathfrak{p}}$ which can be seen as a local foliation of the spacetime $\mathcal{M}$ around the curve $\gamma$ such that any point in $\mathcal{R}$ is described by the Fermi normal coordinates $(\tau, x^1, x^2, x^3)$.

\textit{The system Hamiltonian.} The system Hamiltonian is constructed around the time-like trajectory $\gamma$ employing the Fermi normal coordinates. It is worth mentioning that Ref.~\cite{Perche2022} provides a formal and elegant description of a localized quantum system in a curved spacetime. However, our construction is also related to the formulation reported in Refs.~\cite{Zych2011,Pikovski2015}, where a structured quantum system was considered. We present here only an overview of the calculations, while the details are explicitly given in Sec.~II of the Supplemental Material.  

\begin{figure}
    \centering
    \includegraphics[width=\columnwidth]{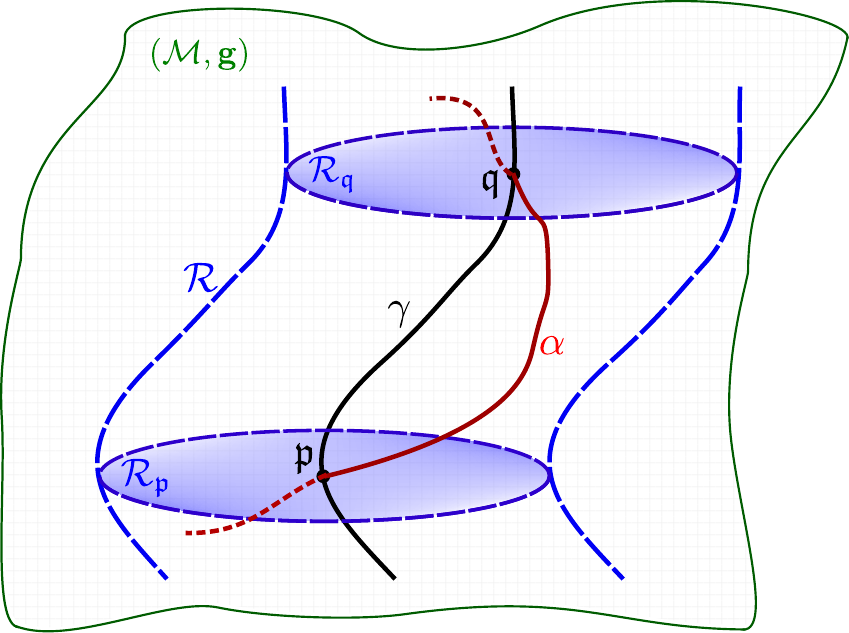}
    \caption{Sketch of the Fermi normal coordinate system. $\mathcal{R}$ is the region where the local rest spaces $\mathcal{R}_{\mathfrak{r}}$, for any $\mathfrak{r}\in\mathcal{M}$, are defined. The Fermi normal coordinate system $(\tau,x^{1},x^{2},x^{3})$ covers each one of these subspaces. The system world-line $\alpha \subset \mathcal{R}$ and the laboratory world-line $\gamma$ are also shown.}
    \label{fig:protocol}
\end{figure}

We consider a particle traveling along the world-line $\alpha$ around the trajectory $\gamma$ of the laboratory frame, i.e., the time-like curve $\alpha$ is contained in the local rest space $\mathcal{R}$ of the curve $\gamma$, where the Fermi normal coordinates are valid. See the sketch in Fig. \ref{fig:protocol}.

Considering a non-relativistic system (the particle's internal and kinetic energies are much smaller than its rest energy), we can write the Hamiltonian to the lowest order as 
\begin{align}
    H(\tau) = & H_{\text{cm}}(\tau) + \mathcal{Z}(\tau) H_{\text{int}} \label{eq:curvhamil}, 
\end{align}
where
\begin{align}
    \mathcal{Z}(\tau) \equiv 1 - \frac{p^2}{2m^2} + a_i(\tau) x^i + \frac{1}{2}R_{\tau i \tau j}(\tau) x^i x^j, \label{eq:timedila}
\end{align}
is directly related to the time dilation factor between the proper time $\tau$ of the laboratory frame and the proper time $t'$ of the system. Besides, $a_{i}(\tau)$ and $R_{\mu \nu \alpha \beta}(\tau)$ are the components, in the Fermi normal coordinates, of the $4$-acceleration and the Riemann curvature tensor, respectively, evaluated at point $\gamma(\tau)$. Finally, $p^{2} = p^{i}p_{i}$ is the square of the momentum of the particle. In addition, $H_{\text{cm}}$ is the Hamiltonian of the center-of-mass, which is given by
\begin{align}
    H_{\text{cm}}(\tau) = m + \frac{p^2}{2m} + m a_i(\tau) x^i + \frac{m}{2}R_{\tau i \tau j}(\tau) x^i x^j. \label{eq:hamilcm}
\end{align}
We can interpret the last two terms of Eq.~\eqref{eq:hamilcm} as perturbations to the flat space-time free Hamiltonian $H_0 = m + p^2/2m$. 

We now have all the necessary ingredients to present our main result.

\textit{Fluctuation theorem.} We start by describing the protocol employed to derive the detailed fluctuation theorem under the TPM scheme for a localized quantum system in a curved spacetime, thus going far beyond Ref.~\cite{Basso2023} in two important ways. First, here we provide a detailed fluctuation theorem, while in Ref.~\cite{Basso2023} just the integral one is presented. Secondly, and most importantly, here we uncover the full role played by the gravitational field in irreversible phenomena, by taking into account the effect of spacetime curvature explicitly. 

Let us first recall that a spacetime $(\mathcal{M}, \mathbf{g})$ is time-orientable if a continuous designation of future-directed and past-directed for time-like vectors can be made over the entire manifold~\cite{Wald}. In our case, we only assume that at least the local rest space $\mathcal{R} \subset \mathcal{M}$ of the curve $\gamma$  is time-orientable. This can be done since we are assuming that the curve $\gamma$ is time-like and describes the world-line of the laboratory frame. Moreover, we assume that the curve $\gamma$ is oriented towards the future. In addition, from the curve $\gamma$ and its local rest space $\mathcal{R}$, it was possible to obtain the Hamiltonian~\eqref{eq:curvhamil} that governs the evolution of our system and therefore defines a notion of time flow in the sense defined by Connes and Rovelli~\cite{Connes1994}. This is crucial for defining local thermal equilibrium states~\cite{Basso2023,Connes1994,Rovelli2011,Rovelli2013}.

We define two protocols, one in the forward direction of time and the other one in the backwards direction. The distinguishability of these processes will be employed as a measure of irreversibility of the forward process~\cite{Crooks1998}. Both processes consist in preparing the system in an equilibrium state, measuring its energy, letting it evolve and measuring its final energy. From the results of these measurements, entropy production can be defined. 

The forward process is defined as follows. As depicted in Fig.~\ref{fig:protocol}, at point $\mathfrak{p} = \gamma(\tau = 0) \in \mathcal{M}$, the curves $\alpha$ and $\gamma$ intersect and the observer (in the laboratory frame) performs the first projective measurement in the energy eigenbasis. To do this, we first suppose that the state of our system is given by $\rho_0 = \ketbra{x_0} \otimes \sigma_{0}$, with $\ket{x_0}$ representing the state of the external degrees of freedom while the internal degrees of freedom are described by the thermal state $\sigma_{0} = e^{-\beta \mathfrak{h}(0)}/Z_0,$ with $\mathfrak{h}(0) = \mathcal{Z}(0) H_{\text{int}}$ and $Z_{0} = \Tr{e^{-\beta \mathfrak{h}(0)}}$ being the initial Hamiltonian of the internal degrees of freedom and the partition function, respectively. $\beta$ is the inverse temperature, which can be properly defined in the present setup~\cite{Basso2023}. The measurement is performed in the eigenbasis of $\mathfrak{h}(0)$ at point $\mathfrak{p} \in \mathcal{M}$, resulting in the energy eigenvalue $\epsilon_{l}^{0}$ with probability $p_{l} = e^{-\beta\epsilon_{l}^{0}}/Z_{0}$. After this measurement, the state of our system is given by $\ket{\Psi_0} = \ket{x_0}\otimes \ket{\epsilon_{l}^{0}}$. Then, as the quantum system travels along its world line $\alpha$, its evolution, with respect to the laboratory frame, is governed by Hamiltonian~\eqref{eq:curvhamil} and can be written as
\begin{eqnarray*}
     \ket{\Psi(\tau)} &=& \mathcal{T} e^{-i\int_{\alpha} \left( H_{\text{cm}}(\tau) + \mathcal{Z}(\tau)H_{\text{int}}\right) \dd \tau} \ket{x_0}\otimes \ket{\epsilon_{l}^{0}} \\
     &=& \mathcal{T}_{\text{cm}} e^{-i\int_{\alpha} H_{\text{cm}} (\tau) \dd\tau} \ket{x_0} \otimes \mathcal{T}_{\text{int}} e^{-i\int_{\alpha} \mathfrak{h}(\tau) \dd\tau} \ket{\epsilon_{l}^{0}},
\end{eqnarray*}
where $\mathcal{T} \equiv \mathcal{T}_{\text{cm}} \otimes \mathcal{T}_{\text{int}}$ is the time-ordering operator and $\mathfrak{h}(\tau) = \mathcal{Z}(\tau) H_{\text{int}}$. The last equality follows from the semiclassical approximation, in which the motion of the quantum particle along its world-line is well-defined. Hence, the internal state evolves accordingly to $U \equiv U(\tau) = \mathcal{T}_{\text{int}} e^{-i\int_{\alpha} \mathfrak{h}(\tau) \dd\tau}$.

At some latter proper time, when the system intersects again the laboratory frame at the point $\mathfrak{q} = \gamma(\tau = T) \in \mathcal{M}$, as depicted in Fig~\ref{fig:protocol}, the last step of the forward process is realized by a projective measurement with respect to the internal Hamiltonian $\mathfrak{h}(T) =\mathcal{Z}(T) H_{\text{int}}$, with  $\mathfrak{h}(T)\ket{\epsilon_{k}^{T}} = \epsilon_{k}^{T}\ket{\epsilon_{k}^{T}}$, $Z_{T} = \Tr{e^{-\beta \mathfrak{h}(T)}}$. Hence, from the definition of work as the stochastic variable $W_{k,l} \equiv \epsilon_{k}^{T} - \epsilon_{l}^{0}$, we can construct the work probability distribution density of the forward process as $P_{\text{fwd}}(W) = \sum_{k,l}p_{k,l}\delta\left[W - W_{k,l}\right]$, where $p_{k,l} = p_{l}p_{k|l}$ is the joint probability of obtaining $\epsilon_{l}^{0}$ in the first measurement and $\epsilon_{k}^{T}$ in the second one. It follows that
\begin{align}
    P_{\text{fwd}}(W)= \sum_{j,k} \delta\left(W - (\epsilon_{k}^{T} - \epsilon_{l}^{0}) \right) \frac{e^{- \beta \epsilon_{l}^{0}}}{Z_0} \abs{\bra{\epsilon_{k}^{0}}U\ket{\epsilon_{l}^{0}}}^2.
\end{align}

In order to define the reverse process, let us remember that, given a time orientation in a portion of the spacetime as the region $\mathcal{R} \subset \mathcal{M}$, we can also define past-directed curves in $\mathcal{R}$~\cite{Wald}. For instance, given that the curves $\gamma$ and $\alpha$ parameterized by $\tau$ in Fig~\ref{fig:protocol} are directed to the future, then we can obtain past-directed curves $\gamma'$ and $\alpha'$ by making $\tau \to - \tau$. The reverse process is defined as follows Ref.~\cite{Campisi}. At point $\mathfrak{q} = \gamma(T) \in \mathcal{M}$, the curves $\alpha$ and $\gamma$ intersect and a first projective measurement in the energy eigenbasis is realized in the laboratory frame. To do this, we first suppose that the state of our system is given by $\rho_{T} = \Theta \ketbra{x_{T}} \otimes \sigma_{T}\Theta^{\dagger}$, where $\Theta$ is the anti-unitary time-reversal operator~\cite{Sakurai}, $\ket{x_{T}}$ is the state of the external degrees of freedom, while the internal degrees of freedom are described by the thermal state $\sigma_{T} = e^{-\beta \mathfrak{h}(\tau)}/Z_{T}$. The next step of the reverse process consists in a measurement in the eigenbasis of $\mathfrak{h}(T)$ at point $\mathfrak{q} \in \mathcal{M}$, resulting in the energy eigenvalue $\epsilon_{k}^{T}$ with probability $p_{k} = e^{-\beta\epsilon_{k}^{T}}/Z_{T}$. 

The time-reversal evolution of internal degrees of freedom is then governed by the micro-reversibility principle, i.e.,  $\Tilde{U} \equiv \Tilde{U}(T - \tau) = \Theta (\mathcal{T}_{\text{int}}e^{-i\int_{\alpha} \mathfrak{h}(\tau) \dd \tau})^{\dagger}\Theta^{\dagger}$, which holds under the assumption that the Hamiltonian~\eqref{eq:curvhamil} is invariant under time-reversal~\cite{Campisi}. When the system intersects the laboratory frame at point $\mathfrak{p} = \gamma(0) \in \mathcal{M}$, the final step of the reverse process is realized by a projective measurement with respect to the internal Hamiltonian $\mathfrak{h}(0)$. Hence, we can compute the work probability distribution density of the reverse process as
\begin{align}
    P_{\text{rev}}(-W)= \sum_{j,k} \delta\left( (\epsilon_{k}^{T} - \epsilon_{l}^{0}) - W \right) \frac{e^{- \beta \epsilon_{k}^{T}}}{Z_{T}} \abs{\bra{\epsilon_{l}^{0}}\Tilde{U}\ket{\epsilon_{k}^{0}}}^2. 
\end{align}
By using the fact that $W = W_{k,l} = \epsilon_{k}^{T} - \epsilon_{l}^{0}$ and $Z_{T}/Z_{0} = e^{- \beta \Delta F}$ where $\Delta F = F_{T} - F_0$ is the difference in the free energy, we obtain
\begin{align}
    \frac{P_{\text{fwd}}(W)}{P_{\text{rev}}(-W)} = e^{\beta(W - \Delta F)}, \label{eq:crooks}
\end{align}

This is the main result of the present paper, the quantum fluctuation theorem in a curved spacetime. It shows that a positive entropy production $\Sigma = W-\Delta F$ will be observed, on average,  every time we are able to tell apart the process from its time reversal.

Moreover, by integrating Eq.~\eqref{eq:crooks} over the probability distributions, we obtain the integral fluctuation theorem
\begin{equation}
\expval{e^{-\beta W}}_{\alpha, \gamma} = e^{-\beta\Delta F}, \label{eq:jar_time} 
\end{equation}
which consists in the relativistic version of the Jarzynski equality. The subscript $\alpha$ in the average above reminds us that the joint probability distribution depends on the path the system follows in the spacetime, while the subscript $\gamma$ reminds us that $\mathcal{Z}(\tau)$ depends on the acceleration of the curve $\gamma$ and the components of the curvature tensor evaluated at the curve $\gamma$. The entropy production as perceived by different observers is discussed in Sec.~IV of the Supplemental Material. Another point is that the final temperature remains identical to the initial one, since it serves merely as a reference state established by the observer at the onset of the process. Therefore, our inquiry revolves around the extent to which the system deviates from this initial equilibrium state during its travel along its path in a curved spacetime. The answer to this question is precisely Eq.~\eqref{eq:jar_time}.

Some comments about particular realizations of our result are needed. First, if $H_{\text{int}}$ depends on $\tau$, the results given by Eqs.~\eqref{eq:crooks} and ~\eqref{eq:jar_time} remain valid with the difference that we also have the contribution of the driven part of the Hamiltonian, thus modifying the entropy production rate. Second, disregarding the internal degrees of freedom, we can also derive Eq.~\eqref{eq:jar_time} for the center-of-mass degrees of freedom by considering localized quantum systems within the local rest space $\mathcal{R}$ of the curve $\gamma$. This is the case of the example discussed in the following. The total Hamiltonian is given by $H(\tau) = H_{\text{cm}}(\tau) = H_0  + m a_i(\tau) x^i + \frac{m}{2}R_{\tau i \tau j}(\tau) x^i x^j,$ where $H_0 = m + \frac{p^2}{2m} + V$ is the non-perturbed Hamiltonian (or the Hamiltonian in flat space-time) with $V$ being the potential energy operator. Therefore, the terms $m a_i(\tau) x^i$ and $\frac{m}{2}R_{\tau i \tau j}(\tau) x^i x^j$ can be treated in the context of quantum mechanical perturbation theory, with the spatial Fermi normal coordinates being position operators and the projective energy measurements are realized with regard to the Hamiltonian $H_{\text{cm}}(\tau)$. 

\textit{The expanding Universe.} We illustrate our results considering a quantum mechanical harmonic oscillator (QHO) in an expanding universe described by the Friedmann-Robertson-Walker (FRW) metric~\cite{Wald}. The detailed calculations and a more in-depth discussion can be found in Sec.~III of the Supplemental Material. 

By taking the worldline of our laboratory frame as the one of the comoving observers ---with the expansion of the universe--- the FRW metric takes, in the Fermi normal coordinate system, the form
\begin{eqnarray}
    \dd s^2 &=& -\Big(1 - \frac{\Ddot{\mathfrak{a}}}{\mathfrak{a}}r^2\Big)\dd\tau^2 \nonumber\\
    &+& \Big[\delta_{ij} - \frac{\dot{\mathfrak{a}}^2}{\mathfrak{a}^2}\Big(\frac{r^2 \delta_{ij} - x_i x_j}{3} \Big) \Big] \dd x^i \dd x^j,
\end{eqnarray}
with $\mathfrak{a}$ being the scale factor.

The system is initially prepared in a thermal state with inverse temperature $\beta$ and associated with the initial Hamiltonian whose spectrum is $\epsilon_k^0 = (k + 1/2)\omega_0$, with $k$ being a non-negative integer. After the first energy measurement, the system is let to evolve under the Hamiltonian
\begin{equation}
   H(\tau) = H_0 + \frac{1}{2}m R_{\tau x \tau x}(\tau) x^2 = H_0 - \frac{m \Ddot{\mathfrak{a}}}{2 \mathfrak{a}} x^2. \label{eq:Huni}
\end{equation}
where the last term of Eq.~\eqref{eq:Huni} plays the role of a time-dependent external potential. Hence, we can interpret the non-stationary spacetime as the external force driving the quantum system out of equilibrium, which is the origin of the entropy production.

Let us restrict ourselves to the de-Sitter solution with the universe dominated by a positive cosmological constant $\Lambda$, which is a model both for the primordial inflationary and the current exponential expansion of our universe, i.e., the matter-energy content of the universe is described by a vacuum with positive energy density which is constant in space and time. In this case we have $\mathfrak{a}(t) = e^{\mathbb{H} t}$ where $\mathbb{H} = \dot{\mathfrak{a}}/\mathfrak{a} = \sqrt{\Lambda/3}$ is the Hubble parameter, and the transition probability (for $k \neq l$) takes the form
\begin{equation}
    p^{\tau}_{k|l} =  4 \Big(\frac{m \mathbb{H}^2}{2}\Big)^2 \frac{\abs{\bra{\epsilon^0_k}x^2\ket{\epsilon^0_l}}^2}{\abs{\epsilon^0_k - \epsilon^0_l}^2} \sin^2\Big(\frac{(k - l)\omega_0 t}{2}\Big). \label{eq:tpho}
\end{equation}
Since this is, in general, different from zero, entropy will be produced by the dynamics unless the Hubble constant is zero. We can interpret this result as a sort of internal friction, that takes information out of the system due to the coupling to the gravitational field. For instance, if we consider initially the system in its ground state, the only transition allowed is for the second excited state and a direct calculation shows that Eq.~\eqref{eq:tpho} give us $p^{\tau}_{2|0} = (\mathbb{H}/\sqrt{2}\omega_0)^4 \sin \omega_0 t$. Given that $\mathbb{H} \approx 10^{-61} t_p^{-1}$, where $t_p = 5.391 \times 10^{-44}s$ is the Planck time, and $\omega_0 \approx 10^{-30} t_p^{-1}$ ($\omega_0 \approx 10^{13}s^{-1}$) for typical molecular vibrational modes, the ratio $\mathbb{H}/\omega_0$ is of order $10^{-31}$, which is very small, but does not vanish. The explicit form of the entropy production is given in Sec.~III of the Supplemental Material.

\textit{Discussion.} In this work, we proved a detailed fluctuation theorem for a localized quantum system living in a general curved spacetime, which reveals how the spacetime curvature can produce entropy.

In order to better understand the role of entropy production due to the curvature of spacetime, let us resort to the gravito-electromagnetic analogy discussed, for instance, in Refs.~\cite{Costa2014, Ruggiero2021} and define the gravito-electric potential as $\phi(\tau) \equiv - \frac{1}{2} R_{\tau i \tau j}(\tau) x^i x^j,$ such that the gravito-electric field (up to linear order in $x^i$) is given by $E_i(\tau) = R_{\tau i \tau j}(\tau) x^j$. The contribution of the gravito-magnetic potential in the gravito-electric field is second order and therefore will not be considered in our analysis. Thus, we can describe the term $\frac{m}{2}R_{\tau i \tau j}(\tau) x^i x^j$ that appears in Eq.~\eqref{eq:hamilcm} as $m E_i(\tau) x^i$, while the term $\frac{1}{2}R_{\tau i \tau j}(\tau) x^i x^j H_{int}$ that appears in $\mathcal{Z}(\tau) H_{\text{int}}$ can be written as $H_{\text{int}} E_i(\tau) x^i$. It is noteworthy the similarity of these two terms with the electric dipole interaction, with both $m$ and $H_{int}$ playing the role of the charge of the gravitational field, which is reasonable since (internal) energy also gravitates in general relativity. Therefore, we can interpret the terms $m E_i(\tau) x^i$ and $ H_{\text{int}} E_i(\tau) x^i$ as the gravitational analogous of a charged quantum system interacting with a time dependent electric field. 

Our main result implies that entropy production is not an invariant quantity defined solely by the system. Rather, it depends on the observer who measures it, once it depends on the world-line of the laboratory in an arbitrary spacetime. This is a robust result that goes in the same direction as those discussed in Refs.~\cite{Wald01,Marolf} regarding the subtleties of defining entropy in a curved spacetime. Specifically, two different families of observers will not agree on the entropy production in general. It is worth remembering that, for comparison, each family of observers has to realize the same protocol, since the measurements in the energy basis are locally performed.

Additionally, our findings establish a deep and fundamental link between the time-orientability of the laboratory frame's world-line $\gamma$ and the production of entropy and, therefore, with the thermodynamic arrow of time. This is the precise meaning of the observer-dependent nature of entropy production. Such orientability is needed in order to obtain the Hamiltonian~\eqref{eq:curvhamil}, which governs the evolution of the quantum system and thereby defines a notion of time flow and thermal equilibrium reference states, as discussed in Refs.~\cite{Connes1994,Rovelli2011,Rovelli2013}. For instance, in the quantum harmonic oscillator in an expanding universe, the curvature drives the quantum system out of equilibrium due to the last term in Eq.~(\ref{eq:Huni}), causing the change in the populations. Therefore, our result shows that the arrow of time is rooted in the causal structure of the spacetime.

Finally, an interesting avenue that it is worth pursuing is the extension to this protocol for quantum fields in curved spacetimes. Another interesting question regards spacetimes with event horizons, in which the information paradox may play a role. 

\begin{acknowledgments}
\textit{Acknowledgments.} This work was supported by the S\~{a}o Paulo Research Foundation (FAPESP), Grant No.~2022/09496-8, by the National Institute for the Science and Technology of Quantum Information (INCT-IQ), Grant No.~465469/2014-0, and by the National Council for Scientific and Technological Development (CNPq), Grants No.~309862/2021-3 and No.~308065/2022-0.
\end{acknowledgments}


\pagebreak
\widetext
\newpage 
\setcounter{secnumdepth}{3}



\begin{center}
\vskip0.5cm
\noindent{\Large \textbf{Supplemental Material}} 
\vskip0.2cm
{\large \textbf{
``Quantum detailed fluctuation theorem in curved spacetimes: \\ The
observer dependent nature of entropy production''
}}
\vskip0.2cm
Marcos L. W. Basso$^1$, Jonas Maziero$^2$, and Lucas C. C\'eleri$^3$
\vskip0.1cm
$^1$\textit{Center for Natural and Human Sciences, Federal University of ABC, States Avenue 5001,\\ Santo Andr\'e, S\~ao Paulo, 09210-580, Brazil} \\
$^2$\textit{Physics Department, Center for Natural and Exact Sciences, Federal University of \\ Santa Maria, Roraima Avenue 1000, 97105-900, Santa Maria, RS, Brazil}\\ 
$^3$\textit{QPequi Group, Institute of Physics, Federal University of Goi\'as, Goi\^ania, GO, 74.690-900, Brazil}
\vskip0.1cm
\end{center}

\setcounter{equation}{0}
\setcounter{figure}{0}
\setcounter{page}{1}
\renewcommand{\thefigure}{S\arabic{figure}}
\renewcommand{\theequation}{S\arabic{equation}}
\renewcommand{\thetable}{S\arabic{table}}

\thispagestyle{empty}

\section{Fermi normal coordinates}
\label{app:fermi}

In the main text, we gave only a bird's eye view of what we call the laboratory coordinate system. For completeness, here we provide the details on how to construct the Fermi normal coordinate system. In this way, some overlap with the main text is inevitable in order to keep reading smooth. 

Let $(\mathcal{M}, \mathbf{g})$ be a four-dimensional spacetime with $\mathcal{M}$ being a differential manifold and $\mathbf{g}$ a Lorentzian metric. Let us also consider a time-like curve $\gamma: I \subset \mathbb{R} \to \mathcal{M}$ parameterized by its proper time $\tau \in I$. This curve represents the world-line of the laboratory frame whose $4$-velocity $u^{\mu}$ fulfills $u_{\mu} u^{\mu} = - 1$. 

To construct the Fermi normal coordinates, we start by defining the proper time $\tau$ of the curve $\gamma$ as the time component of such a coordinate system. Then we define an orthonormal basis $e_{a}^{\ \mu}$ at a point $\mathfrak{p} \equiv \gamma(\tau = 0) \in \mathcal{M}$, where $a = 0, 1, 2, 3$ labels the four basis vectors and $e_0^{\ \mu}$ is identified with the tangent vector $u^{\mu}$. Thus, at point $\mathfrak{p}$, we have
\begin{align}
    g_{\mu \nu} e_{a}^{\ \mu} e_{b}^{\ \nu} =  \eta_{a b},
\end{align}
where $\eta_{a b} = \text{diag}(-1,1,1,1)$ is the Minkowski metric. The next step consists in extending this frame along the trajectory $\gamma$ such that the basis remains orthogonal. To do this, let us first note that the parallel transport along the curve $\gamma$ does not guarantee that the vectors in the set $\{e_{a}^{\ \mu}\}_{a = 0}^3$ remain orthogonal to each other, except in the case of $\gamma$ being a geodesic. Therefore, in order to extend the orthonormal frame defined by the set $\{e_{a}^{\ \mu}\}_{a = 0}^3$ such that the set remains orthonormal along the curve $\gamma$, it is necessary to transport the vectors $e_{a}^{\ \mu}$ via the Fermi-Walker transport, which is defined by the following set of differential equations~\cite{hawkingellis}:
\begin{align}
    \frac{\textrm{D}_F}{\dd \tau} (e_{a})^{\mu} \equiv \frac{\textrm{D}}{\dd \tau} (e_{a})^{\mu} + 2 a^{[\mu}u^{\nu]}(e_a)_{\nu} = 0, \label{eq:fermit}
\end{align}
where $\textrm{D}/\dd \tau \equiv u^{\mu} \nabla_{\mu}$ is the covariant derivative along the curve $\gamma$, $a^{\mu} = (\textrm{D}/\dd \tau)\, u^{\mu} = u^{\nu} \nabla_{\nu} u^{\mu}$ is the $4$-acceleration of the curve $\gamma$ while $2 a^{[\mu}u^{\nu]} \equiv a^{\mu} u^{\nu} - a^{\nu} u^{\mu}$. Therefore, if Eq.~\eqref{eq:fermit} holds we say that the vectors $e_{a}^{\ \mu}$ are Fermi transported. It is worth noticing that the Fermi-Walker transport obeys the following properties: (i) it reduces to the parallel transport when the curve $\gamma$ is a geodesic; (ii) the tangent vector $u^{\mu}$ is always Fermi transported along the curve $\gamma$; (iii) if any two vectors $v^{\mu}$ and $w^{\mu}$ are Fermi transported along the curve $\gamma$, then the inner product $v_{\mu}w^{\mu}$ is constant along $\gamma$. This is the laboratory frame employed in the main text.

In order to define the space-like Fermi normal coordinates $(x^1, x^2, x^3)$, it is necessary to consider the normal neighborhood of the point $\mathfrak{p} = \gamma(\tau = 0)$, which is set of all points that can be connected to $\mathfrak{p}$ by a single geodesic and it is denoted by $\mathcal{U}_{\mathfrak{p}}$. Then we can define the local rest space $\mathcal{R}_{\mathfrak{p}} \subset \mathcal{U}_{\mathfrak{p}}$ of the point $\mathfrak{p}$ as the set of points spanned by all geodesics that start from $\mathfrak{p}$ with tangent vector orthogonal to $u^{\mu}$. Therefore, we can ascribe the coordinates $(\tau = 0, x^1, x^2, x^3)$ to any point $\mathfrak{r} \in \mathcal{R}_{\mathfrak{p}}$ through the exponential map such that $\mathfrak{r} = \exp_{\mathfrak{p}}(x^a e_a)$, where $\exp_{\mathfrak{p}}: T_{\mathfrak{p}}(\mathcal{M}) \to \mathcal{M}$ is the exponential map at the point $\mathfrak{p}$, $T_{\mathfrak{p}}(\mathcal{M})$ is the tangent space of $\mathfrak{p}$ and $e_a \in T_{\mathfrak{p}}(\mathcal{M})$. Finally, the local rest space of the curve $\gamma$ is defined as $\mathcal{R} \equiv \cup_{\mathfrak{p} \in \gamma} \mathcal{R}_{\mathfrak{p}} $, which can be seen as a local foliation of the spacetime $\mathcal{M}$ around the curve $\gamma$ such that any point in $\mathcal{R}$ is described by the Fermi normal coordinates $(\tau, x^1, x^2, x^3)$. As a consequence of the definition of the Fermi normal coordinates and the decomposition of the metric along the curve $\gamma$ as $g_{\mu \nu} = - u_{\mu} u_{\nu} + \delta_{i j} e^{i}_{\ \mu} e^{j}_{\ \nu}$, the spatial distance of a point $\mathfrak{r} \in \mathcal{R}$ to the curve $\gamma$ is given by $r^2 = \delta_{i j}x^i x^j$ \cite{Poisson2011}.

Finally, the Fermi normal coordinates $(\tau, x^1, x^2, x^3)$ allow us to express the components of the metric around the curve $\gamma$ as
\begin{align}
 & g_{\tau \tau} = - (1 + a_i(\tau) x^i)^2 - R_{\tau i \tau j}(\tau) x^i x^j + \mathcal{O}(r^3), \nonumber \\
 & g_{\tau i} = - \frac{2}{3} R_{\tau jik}(\tau) x^j x^k + \mathcal{O}(r^3), \label{app:eq:metricfnc}\\
 & g_{i j} = \delta_{ij} - \frac{1}{3} R_{ikjl}(\tau)x^k x^l + \mathcal{O}(r^3), \nonumber
\end{align}
where $a^{\mu}(\tau)$ and $R_{\mu \nu \alpha \beta}(\tau)$ represent, respectively, the $4$-acceleration and components of the Riemann curvature tensor in the Fermi normal coordinates evaluated at the point $\gamma(\tau)$.  If the curve $\gamma$ is a geodesic, the Fermi-Walker transport reduces to the parallel transport and the Fermi normal coordinates reduce to the Riemann normal coordinates. 

\section{Hamiltonian dynamics of a localized quantum system in curved spacetimes}
\label{app:hamil}
We start by giving the classical description of the Hamiltonian of a particle with some internal structure, which we later quantize.

Let us consider a particle with some internal structure traveling along its world-line $\alpha$ around the trajectory $\gamma$ of the laboratory frame, i.e., the time-like curve $\alpha$ is contained in the local rest space $\mathcal{R}$ of the curve $\gamma$, where the Fermi normal coordinates $(\tau, x^1, x^2 , x^3)$ are valid. See the sketch in Fig.~1 of the main text. In this coordinate system, we describe the $4$-momentum of the particle along the world-line $\alpha$ as $p^{\mu}$. Whereas, in the particle rest frame, denoted by primed coordinates $x^{\mu'}$, it is easy to see that $p^{j'} =  (\partial x^{j'}/\partial x^{\mu}) p^{\mu} = 0$ (with $j$ labeling only the spatial coordinates), which implies that the total energy, as measured by the commoving observer, is given by $p_{t'}$ ($x^{0'} \equiv t'$). It comprises not only the energy stemming from the rest mass of the system but also any binding or kinetic energies of the internal degrees of freedom and thus also the particle's internal Hamiltonian $H_{\text{int}}$. Therefore
\begin{align}
    p_{t'} = m + H_{\text{int}} \equiv H_{\text{rest}}. \label{eq:Hrest}
\end{align}

In contrast, $p^{\tau}$ describes the dynamics of the particle with respect to the laboratory frame associated with the Fermi normal coordinate system, which includes the energy of both internal and external degrees of freedom. Therefore, it constitutes the total Hamiltonian of the system relative to $(\tau, x^1, x^2 , x^3)$ and will be denoted by $H = p_{\tau}$. Given that $p_{\mu} p^{\mu} = p_{\mu'}p^{\mu'}$ is a coordinate invariant quantity, we have the following relation between $H$ and $H_{\text{rest}}$:
\begin{align}
    H = \sqrt{\frac{g^{t't'}H^2_{\text{rest}} - p_j p^j}{g^{\tau \tau}}}.
\end{align}
Taking the component $x^{0'} = t'$ associated with the commoving observer to be the proper time along the particle's world line implies that $g^{t't'} = -1$, and therefore
\begin{align}
    H & = \sqrt{-\frac{H^2_{\text{rest}} + p_j p^j}{g^{\tau \tau}}}  = z(\tau) \sqrt{H^2_{\text{rest}} + p_j p^j},
\end{align}
where we notice that $z(\tau) = (- g^{\tau \tau})^{-1/2} = \abs{g^{\tau \tau}}^{-1/2}$ is the red-shift factor.    
 
Now, in the non-relativistic limit, we can expand the red-shift factor as
\begin{align}
    z(\tau) \approx 1 + a_i(\tau) x^i + \frac{1}{2}R_{\tau i \tau j}(\tau) x^i x^j,  \label{eq:redshifapp}
\end{align}
since the main contribution of the red-shift factor stems from the terms $H_{\text{rest}} a_i(\tau) x^i$ and $\frac{H_{\text{rest}}}{2}R_{\tau i \tau j}(\tau) x^i x^j$, as discussed in Ref.~\cite{Perche2022}. It is worth mentioning that, in Ref.~\cite{Perche2022}, the authors only consider the contribution due to the rest mass such that $H_{\text{rest}} = m$. In this limit, we also have $\sqrt{H^2_{\text{rest}} + p_j p^j} \approx H_{\text{rest}} + \frac{p^2}{2  H_{\text{rest}}}$ where $p^2 \equiv p_j p^j$. Therefore, using Eq.~\eqref{eq:Hrest}, at the lowest order, we have
\begin{align}
    H(\tau) = & H_{\text{cm}}(\tau) + \mathcal{Z}(\tau) H_{\text{int}} \label{app:eq:curvhamil}, 
\end{align}
where $\mathcal{Z}(\tau)$ is given in Eq.~(2) of the main text, while $H_{\text{cm}}$ stands for the Hamiltonian of the center-of-mass (Eq.~(3) of the main text).

The quantity $\mathcal{Z}(\tau)$ is directly related to the time dilation factor between the proper time $\tau$ of the laboratory frame and the proper time $t'$ of the system of interest (or the commoving observer). To see this, we remember that $-dt'^2 = -g_{\mu \nu} dx^{\mu} dx^{\nu}$, where $x^{\mu}$ are the Fermi normal coordinates, thus implying
\begin{equation}
    \frac{d t'}{d \tau} = \sqrt{-g_{\tau \tau} - g_{ij} \frac{d x^i}{d \tau}\frac{d x^j}{d \tau}} \approx 1 - \frac{p^2}{2m^2} + a_i(\tau) x^i + \frac{1}{2}R_{\tau i \tau j}(\tau) x^i x^j = \mathcal{Z}(\tau).
\end{equation}

Now, if the system in question is described by quantum mechanics with some internal degree of freedom, we can canonically quantize this classical framework by considering that this system is associated with the Hilbert space $\mathcal{H_{\text{cm}}} \otimes \mathcal{H}_{\text{int}}$, encompassing both the center-of-mass and internal degrees of freedom. The total Hamiltonian $H$, as defined in Eq.~\eqref{app:eq:curvhamil}, becomes a Hermitian operator acting in this composite Hilbert space. Here, $H_{\text{cm}}$ governs the particle's external dynamics, while the additional term in Eq.~\eqref{app:eq:curvhamil} addresses its internal dynamics.

With this construction, we can write the Schr\"odinger's equation that governs the unitary evolution of our quantum system, whose state is denoted by $\ket{\psi} \in \mathcal{H}^{(\tau)}_{\text{cm}} \otimes \mathcal{H}_{\text{int}}$, where $\mathcal{H}^{(\tau)}_{\text{cm}} \simeq L^2(\mathcal{R}_{\tau})$ with $L^2(\mathcal{R}_{\tau})$ being the space of square integrable functions in the rest space $\mathcal{R}_{\tau}$ and $\tau$ is the proper time of the world-line of the laboratory frame $\gamma$ which we use here to label a given point $\mathfrak{p}$ of such world-line. Hence, the Schr\"odinger's equation reads
\begin{align}
    i \frac{d}{d \tau} \ket{\psi} = H(\tau) \ket{\psi}, \label{eq:schro}
\end{align}
with $H(\tau)$ being the Hamiltonian operator given in the main text. From Eq.~\eqref{eq:schro} we can define the global unitary evolution operator $U$ by $U(\tau) \ket{\psi(\tau = 0)} = \ket{\psi(\tau)}$ such that $U:\mathcal{H}^{(0)}_{\text{cm}} \otimes \mathcal{H}_{\text{int}} \to \mathcal{H}^{(\tau)}_{\text{cm}} \otimes \mathcal{H}_{\text{int}}$ (for more details see Ref.~\cite{Perche2022}).

Moreover, let us notice that, in our case, the Hamiltonian obtained in this procedure is self-adjoint since it is quadratic in the momentum $p$. However, in a more general context, this may not be true and the procedure does not necessarily lead to a self-adjoint Hamiltonian. Therefore, in order to obtain a self-adjoint Hamiltonian, one can postulate a new Hamiltonian via symmetrization as $\mathcal{H} = \frac{1}{2}(z(\tau) H + \text{t.c.})$ with $\text{t.c.}$ being the transpose conjugate and $H = H_{\text{cm}} + H_{\text{int}}$. Then we expand $z(\tau)$ to obtain the first order corrections, as discussed in Ref.~\cite{Perche2022}. 

Another point worth mentioning is that the red shift factor $z(\tau)$ is a function of the space-like Fermi coordinates, while the time dilation factor $\mathcal{Z}(\tau)$ is a function of the space-like Fermi coordinates and the momentum of the center of mass, as one can see from Eqs.~\eqref{eq:redshifapp} and Eq.~(3) of the main text, respectively. This fact implies that the quantization procedure promotes the space dependence of $z(\tau)$ to a function of the position operator and $\mathcal{Z}(\tau)$ to a function of the position and momentum operators that act on the composite Hilbert space $\mathcal{H}_{\text{cm}} \otimes \mathcal{H}_{\text{int}}$. 

\section{The expanding Universe} 
\label{app:universe}

\begin{figure}
    \centering
    \includegraphics[scale=0.6]{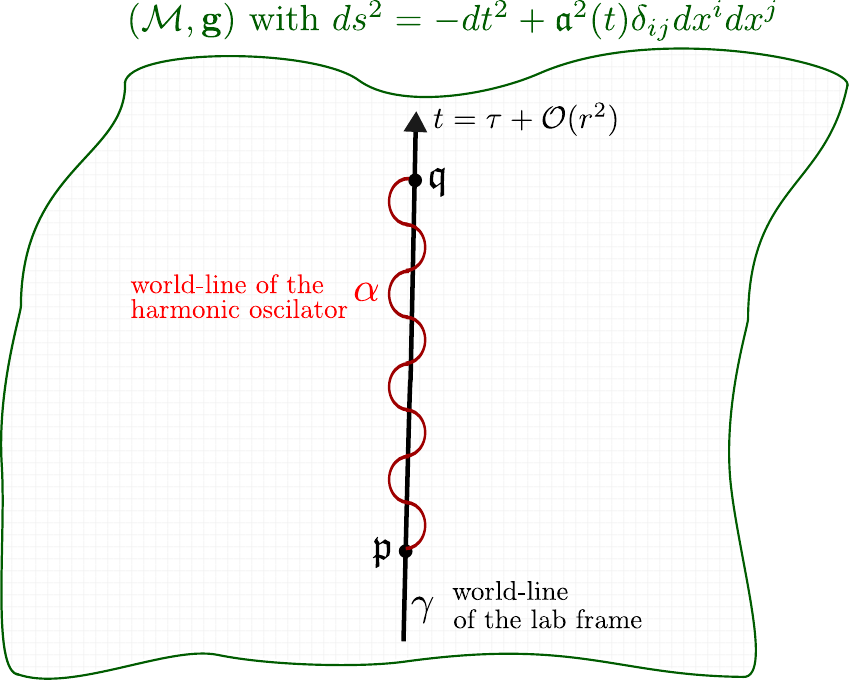}
    \caption{Harmonic oscillator in an expanding universe.}
    \label{fig:universe}
\end{figure}

We illustrate our results by considering a quantum mechanical harmonic oscillator (QHO) in an expanding universe as depicted in Fig.~\ref{fig:universe}. This example was chosen to show that our protocol also works for the center-of-mass degrees of freedom, disregarding the internal degrees of freedom. Alternatively, we could also consider that the QHO can model the vibrational modes of a diatomic molecule, which plays the role of the internal degrees of freedom, whose center of mass is treated classically and follows the same world-line of the laboratory frame $\gamma$.

More specifically, let us first consider an isotropic and homogeneous expanding universe with zero spatial curvature \cite{Wald}. In the Friedmann-Robertson-Walker (FRW) coordinates, the metric can be written as
\begin{align}
    ds^2 = -dt^2 + \mathfrak{a}^2(t)\Big(dX^2 + dY^2 + dZ^2\Big),
\end{align}
where $\mathfrak{a}(t)$ is the scale factor. The world-line of our laboratory frame will coincide with the world-line of the commoving observers with the expansion of the universe which follows the geodesic $\gamma$ given by $\tau = \lambda$ and $X = Y = Z = 0$, implying that the acceleration of our laboratory frame is zero. In the Fermi normal coordinates $\{\tau, x, y, z\}$, the metric on the curve $\gamma$ is the Minkowski metric $\eta_{a b} = \text{diag}(-1,1,1,1)$ and the Fermi propagated orthonormal tetrad is given by
\begin{align}
     e^F_0 = (1, 0, 0, 0), \ \ \ e^F_1 = (0, 1, 0, 0), \ \ \ e^F_2 = (0, 0, 1, 0), \ \ \ e^F_3 = (0, 0, 0, 1), 
\end{align}
which in this case is parallel propagated given that $\gamma$ is a geodesic. The transformation from FRW coordinates $\{t, X, Y, Z\}$ to the Fermi normal coordinates $\{\tau, x , y, z\}$ is given by \cite{Cooperstock1998} 
\begin{align}
    & t = \tau - \frac{\dot{\mathfrak{a}}}{2 \mathfrak{a}}r^2 + \mathcal{O}(r^4),\ \ \ X^i = \frac{x^i}{\mathfrak{a}} \Big(1 + \frac{\dot{\mathfrak{a}}^2}{3 \mathfrak{a}^2} r^2\Big) + \mathcal{O}(r^4), 
\end{align}
where $r \equiv \delta_{ij}x^i x^j$ and $\dot{\mathfrak{a}} = d \mathfrak{a}/ d t$. It is worth noticing that, at lowest order, we have $t = \tau$ and $X^i = x^i/\mathfrak{a}$. Therefore, in the Fermi normal coordinates, the metric takes the form
\begin{align}
    ds^2 = -\Big(1 - \frac{\Ddot{\mathfrak{a}}}{\mathfrak{a}}r^2\Big)d\tau^2 + \Big[\delta_{ij} - \frac{\dot{\mathfrak{a}}^2}{\mathfrak{a}^2}\Big(\frac{r^2 \delta_{ij} - x_i x_j}{3} \Big) \Big] dx^i dx^j.
\end{align}
By comparing with Eq.~\eqref{app:eq:metricfnc}, it follows that the components of the curvature in the Fermi normal coordinates are given by \cite{Cooperstock1998}
\begin{align}
    & R_{\tau x \tau x} = R_{\tau y \tau y} = R_{\tau z \tau z} = - \frac{\Ddot{\mathfrak{a}}}{\mathfrak{a}}, \ \ \ \ R_{xyxy} = R_{xzxz} = R_{yzyz} = \frac{\dot{\mathfrak{a}}^2}{\mathfrak{a}^2}.
\end{align}

Given that the spacetime scenario is set, let us consider a one-dimensional QHO whose energy eigenvalues are given by $\epsilon_n^0 = (n + 1/2)\omega_0$ regarding the unperturbed Hamiltonian $H_0 = \frac{p^2}{2m} + \frac{1}{2}m \omega_0^2 x^2$. Here, we ignore the rest mass energy term, since the only effect is a shift on the energy levels. Let us also notice that, in this case, the Fermi normal coordinate $x$ is a position operator on the Hilbert space of the system.

Our protocol starts with the quantum system in thermal equilibrium at some point $\mathfrak{p} = \gamma(0)$ of the world-line of the laboratory frame, where a projective measurement in the eigenbasis $\ket{\epsilon^0_l}$ of $H_0$ is realized. The probability of measuring the eigenvalue $\epsilon^0_l$ is given by 
\begin{equation}
 p_l = e^{-\beta \epsilon^0_l}/Z_0,  \label{eq:pl} 
\end{equation}
with $Z_0 = 2/\sinh(\frac{1}{2}\beta \omega_0)$. The next step consists in letting the quantum system evolve with the Hamiltonian, according to the laboratory frame, given by
\begin{align}
    H(\tau) = H_0 + \frac{1}{2}m R_{\tau x \tau x}(\tau) x^2 = H_0 - \frac{m \Ddot{\mathfrak{a}}}{2 \mathfrak{a}} x^2. \label{Htau} 
\end{align}
From Eq.~(9) of the main text, we can see that a non-stationary gravitational field influences the non-relativistic quantum system such that the frequencies become time-dependent, i.e.,
\begin{align}
\omega(\tau) = \sqrt{\omega^2_0 - \frac{\Ddot{\mathfrak{a}}}{\mathfrak{a}}}. \label{eq:freq}    
\end{align}
Hence, the energy of such oscillator is not conserved and, consequently, there are non-null transition probabilities between the initial and the final energy states.
 
Thus, we can use time-dependent perturbation theory or the interaction picture~\cite{Sakurai} in order to obtain the transition probabilities. In the interaction picture, we have $i \partial_{\tau} \ket{\psi}_I = V_I(\tau) \ket{\psi}_I$ with $V_I(\tau) = e^{i H_0 \tau} V(\tau) e^{-i H_0 \tau}$ and $\ket{\psi}_I = e^{i H_0 \tau} \ket{\psi}$. 
By expanding $\ket{\psi}_I = \sum_k c_k(\tau) \ket{\epsilon_k^0}$, it gives us
\begin{align}
    c_k(\tau) = - i \sum_l \int^{\tau}_0  \bra{\epsilon^0_k}V(\tau')\ket{\epsilon^0_l} e^{i(k - l) \tau'} c_l(\tau') d \tau'.
\end{align}
Given that the initial state of our system is $\ket{\epsilon^0_l}$, we obtain
\begin{align}
c_k(\tau) = - \frac{i m }{2} \bra{\epsilon^0_k}x^2\ket{\epsilon^0_l} f(\tau),   
\end{align}
where
\begin{align}
  \bra{\epsilon^0_k}x^2\ket{\epsilon^0_l}  =  \frac{1}{2m\omega_0}  \Big(\sqrt{l(l-1)}\delta_{k, l-2} + (2l +1) \delta_{k,l} +  \sqrt{(l+1)(l+2)}\delta_{k, l+2}\Big)
\end{align}
and 
\begin{align}
    f(\tau) = \int_{0}^{\tau} R_{\tau x \tau x} (\tau') e^{i(k-l) \omega_0 \tau' } d \tau' = -  \int_{0}^{\tau} \frac{\Ddot{\mathfrak{a}}}{\mathfrak{a}} e^{i(k-l) \omega_0 \tau' } d \tau'. 
\end{align}
Above we used the fact that $x = \sqrt{\frac{1}{2m\omega_0}}(A + A^{\dagger})$ with $A$ and $A^{\dagger}$ being the annihilation and the creation operators, respectively. The transition probability to a state $\ket{\epsilon^T_k}$, with $k \neq l$, is given by 
\begin{align}
p_{k|l} = \abs{c_{k \neq l}(T)}^2 \label{eq:pk|l}.    
\end{align}
It is interesting to notice that the transition probability depends on the curvature of the expanding universe. From Eqs.~\eqref{eq:pl} and~\eqref{eq:pk|l}, it is straightforward to compute the work distribution of the forward process, i.e.,
\begin{align}
    P_{\text{fwd}}(W)= \sum_{k,l} \delta\left(W - (\epsilon_{k}^{T} - \epsilon_{l}^{0}) \right) \frac{e^{- \beta \epsilon_{l}^{0}}}{Z_0} \abs{c_{k \neq l}(T)}^2.
\end{align}

For the reverse process, the protocol starts with a measurement in the eigenbasis of $\mathfrak{h}(T)$ at point $\mathfrak{q} \in \mathcal{M}$, resulting in the energy eigenvalue $\epsilon_{k}^{T}$ with probability 
\begin{align}
p_{k} = e^{-\beta\epsilon_{k}^{T}}/Z_{T},    
\end{align}
where $Z_T = 2/\sinh(\frac{1}{2}\beta \omega_T)$. The time-reversal evolution of internal degrees of freedom is implemented by $\Tilde{U} \equiv \Tilde{U}(T - \tau) = \Theta (\mathcal{T}_{\text{int}}e^{-i\int_{\alpha} \mathfrak{h}(\tau) d \tau})^{\dagger}\Theta^{\dagger}$. Here, it is important to notice that the Hamiltonian given by Eq.~\eqref{Htau} is invariant under time-reversal for scale factors used in the literature to describe relevant expanding spacetime models~\cite{Cooperstock1998}, such as $\mathfrak{a}(t) = \alpha t^n$ and $\mathfrak{a}(t) = e^{\mathbb{H} t}$. Hence, the reverse transition probability of obtaining the state $\ket{\epsilon^0_l}$ is given by
\begin{align}
p_{l|k} =  \abs{c_{k \neq l}(T)}^2 \label{eq:pl|k}.    
\end{align}
A straightforward calculation shows that the work distribution of the reverse process is
\begin{align}
    P_{\text{rev}}(-W)= \sum_{k,l} \delta\left( (\epsilon_{k}^{T} - \epsilon_{l}^{0}) - W \right) \frac{e^{- \beta \epsilon_{k}^{T}}}{Z_{T}} \abs{c_{k \neq l}(T)}^2, 
\end{align}
which allows us to confirm the detailed fluctuation theorem obtained in the main text.

In particular, let us consider the  de-Sitter solution with the universe dominated only by a positive cosmological constant $\Lambda$. By writing the Einstein field equation as $R_{\mu \nu} - \frac{1}{2}R g_{\mu \nu} = - \Lambda g_{\mu \nu}$, the energy-momentum tensor can be written as $T_{\mu \nu} = - \frac{\Lambda}{8 \pi}g_{\mu \nu}$, which corresponds to the energy-momentum tensor of a perfect fluid such that $p_{\Lambda} = - \rho_{\Lambda} = - \frac{\Lambda}{8 \pi}$, where $p_{\Lambda}$ is the isotropic pressure and $\rho_{\Lambda}$ is the positive energy density. In this case, we have that $\mathfrak{a}(t) = e^{\mathbb{H} t}$ where $\mathbb{H} = \frac{\dot{\mathfrak{a}}}{\mathfrak{a}} = \sqrt{\Lambda/3}$ is the Hubble parameter. Hence $\frac{\Ddot{\mathfrak{a}}}{\mathfrak{a}} = \mathbb{H}^2$ and the perturbation corresponds to a constant external potential such that $V(\tau) = - \frac{1}{2}m \mathbb{H}^2 x^2$ for $\tau > 0$, which implies that the transition probability for $k \neq l$ is the one given in the main text, being invariant under time reversal.

Moreover, we can calculate the dissipated work, $W_{\text{diss}} = \expval{W} - \Delta F$, in this irreversible process, where $\expval{W} = \Tr \{H(\tau) \rho_{\tau}\} - \Tr \{H_0 \rho_{0}\}$ is the average work of this process with $\rho_0 = e^{-\beta H_{0}}/Z_{0}$ being the initial thermal state while $\rho_{\tau} = e^{-\beta H_{\tau}}/Z_{\tau}$ is the final thermal state. Therefore, the average entropy production associated with the irreversible work is given by $\Sigma = \beta W_{\text{diss}}$. A straightforward calculation shows that
\begin{align}
    \Sigma  
    & = \frac{\beta \omega_{\tau}}{ e^{\beta \omega_{\tau}} - 1} - \frac{\beta \omega_{0}}{e^{\beta \omega_{0}} - 1} - \ln\left(\frac{1 - e^{-\beta \omega_{\tau}}}{1 - e^{-\beta \omega_{0}}} \right) \label{eq:Sprod}
\end{align}
where  $\omega_{\tau}$ is given by Eq.~\eqref{eq:freq}.

\section{Entropy production for different observers}

In this section, we discuss in more detail the entropy production as perceived by different observers. 
Let us recall that a unit time-like congruence of curves can be interpreted as a family of world-lines of certain observers in a given spacetime that never intersects each other~\cite{hawkingellis}. Considering the scenario outlined in the main text, we analyze a particle traveling along a well-defined world-line $\alpha$ near the trajectory $\gamma$ associated with the laboratory frame, denoted here by $\mathcal{L}$. Now, let us introduce another observer, denoted by $\mathcal{O}$, who moves along a distinct time-like curve $\beta$. Let us suppose that this curve intersects both the curves $\gamma$ and $\alpha$ at the same two points $\mathfrak{p}$ and $\mathfrak{q}$, which correspond to the location where the initial and final measurements in the energy basis are realized.
Thus the second observer $\mathcal{O}$ belongs to a different time-like congruence of curves than the laboratory frame $\mathcal{L}$.

In the reference frame of $\mathcal{O}$, it is also possible to construct a Fermi normal coordinate system $(\tau', y^1, y^2, y^3)$ around the time-like curve $\beta$, which is distinct from the Fermi normal coordinates $(\tau, x^1, x^2, x^3)$ of the laboratory frame $\mathcal{L}$, with the transformation between the different coordinate systems defined in the regions where the local rest spaces of the observers $\mathcal{L}$ and $\mathcal{O}$ overlap. By using the same procedure outlined in Sec.~\ref{app:fermi}, we can build the local rest space of the observer $\mathcal{O}$ and express the spacetime metric in terms of Fermi coordinates established in $\mathcal{O}$. Moreover, following the same steps of Sec.~\ref{app:hamil}, it is possible to obtain the system Hamiltonian and describe the dynamics of the system from the perspective of the observer $\mathcal{O}$. The main point is that, for a given spacetime, the two observers will have different local rest spaces. Within the local rest space of each observer, the Hilbert space of the quantum system can be constructed and all the relevant quantities can be defined. 

In this new reference frame, it is easy to see that the Hamiltonian of the system is given by
\begin{align}
    H(\tau') = & H_{\text{cm}}(\tau') + \mathcal{Z}(\tau') H_{\text{int}} \label{eq:curvhamilO}, 
\end{align}
where
\begin{align}
    & H_{\text{cm}}(\tau') = m + \frac{p'^2}{2m} + m a_i(\tau') y^i + \frac{m}{2}R_{\tau' i \tau' j}(\tau') y^i y^j \label{eq:hamilcmO}, \\
    & \mathcal{Z}(\tau') \equiv 1 - \frac{p'^2}{2m^2} + a_i(\tau') y^i + \frac{1}{2}R_{\tau' i \tau' j}(\tau') y^i y^j, \label{eq:timedilaO}
\end{align}
with $a_{i}(\tau')$ and $R_{\mu \nu \alpha \beta}(\tau')$ are now the components of the $4$-acceleration and the Riemann curvature tensor, respectively, evaluated at point $\beta(\tau')$ in the Fermi normal coordinates of $\mathcal{O}$. Moreover, $p'^{2} = p^{i}p_{i}$ is the square of the momentum of the particle in the new reference frame.  On the other hand, in the laboratory frame $\mathcal{L}$, this quantities are evaluated at the point $\gamma(\tau)$. It is worth mentioning that this is different from just a coordinate transformation, where this quantities are being evaluated at the same point in different coordinate systems. 

For simplicity, let us consider that this observer $\mathcal{O}$ is commoving with the particle, meaning $\beta = \alpha$ and the internal degree of freedom is a spin, which implies that $y^i$ only acts on the external degrees of freedom that can be treated classically. In the reference frame of $\mathcal{O}$, the particle remains at rest and $y^i = 0$ leading to $H_{\text{cm}}(\tau') = m$ and $\mathcal{Z}(\tau') = 1$. As a result, the internal Hamiltonian simplifies to $\mathfrak{h}(\tau') = \mathcal{Z}(\tau') H_{\text{int}} = H_{\text{int}}$. This internal Hamiltonian differs from the one described in the laboratory frame $\mathcal{L}$, except at the two intersection points $\mathfrak{p}$ and $\mathfrak{q}$. Consequently, the dynamics of the system as perceived by the observer $\mathcal{O}$ will be different of the dynamics described by the observer $\mathcal{L}$ and the average entropy production calculated by each observer will not be the same. The subscript $\gamma$ in Eq.~(7) of the main text remember us exactly of this fact.  

Let us notice that, since the curves of the two observers intersect at the beginning and at the end of the protocol, the observers will agree on the definition of the thermal states and their corresponding temperatures. The same thing happens for the reversed protocol.

As a last comment, it is worth mentioning that, in a flat spacetime, the components of the curvature tensor vanish at every point and in every reference frame. Furthermore, in such spacetime, there exists a privileged family of inertial observers. For these observers, the $4$-acceleration along their world-lines is zero. Consequently, the original fluctuation theorem is naturally recovered.


\end{document}